\documentclass[oupdraft]{bio}
\usepackage{url, algorithm2e,  multirow}

\history{Received August 1, 2010;
revised October 1, 2010;
accepted for publication November 1, 2010}

\begin{document}

\title{Missing data interpolation in integrative multi-cohort analysis with disparate covariate information}

% List of authors, with corresponding author marked by asterisk
\author{EKATERINA SMIRNOVA$^{1\ast}$, 
YONGQI ZHONG$^{2}$, RASHA ALSAADAWI$^{1}$, XU NING$^{2}$, AMII KRESS$^{2}$, JORDAN KUIPER$^{2}$, MINGYU ZHANG$^{2}$, 
KRISTEN LYALL$^{3}$, SHEENAS MARTENIES$^{4}$, AKRAM ALSHAWABKEH$^{5}$, CATHERINE BULKA$^{6}$, CARLOS CAMARGO$^{7}$, JAEUN CHOI$^{8}$, ELENA COLICINO$^{9}$, ANNE DUNLOP$^{10}$, MICHAEL ELLIOTT$^{11}$, ASSIAMIRA FERRARA$^{12}$, TEBEB GEBRESTADIK$^{13}$, JIANG GUI$^{14}$, KYLIE HARRALL$^{15}$, TINA HARTERT$^{13}$, BARRY LESTER$^{16}$, ANDREW MANIGAULT$^{16}$, JUSTIN MANJOURIDES$^{5}$, YU NI$^{17}$, ROSALIND WRIGHT$^{9}$, ROBERT WRIGHT$^{9}$, KATHERINE ZIEGLER$^{18,19}$, and BRYAN LAU$^{2}$, on behalf of program collaborators for Environmental influences on Child Health Outcomes*\\[3pt]
% Author addresses
\textit{$^{1}$Department of Biostatistics, School of Medicine,
Virginia Commonwealth University,
MCV Box 980032, Richmond, VA 23298-0032}
{$^{\ast}$ekaterina.smirnova@vcuhealth.org}\\
\textit{$^{2}$Department of Epidemiology,
Bloomberg School of Public Health,
Johns Hopkins University,
615 N Wolfe St,
Baltimore, MD 21205
USA}
\\[1pt]
\textit{$^{3}$AJ Drexel Autism Institute,
Drexel University,
Philadelphia, PA, USA}
\\[1pt]
\textit{$^{4}$Department of Kinesiology and Community Health, 
College of Applied Health Sciences,
University of Illinois Urbana-Champaign,
Champaign, IL, USA}
\\[1pt]
\textit{$^{5}$Department of Civil and Environmental Engineering,
Northeastern University College of Engineering,
Boston, MA, USA}
\\[1pt]
\textit{$^{6}$University of South Florida College of Public Health,
Tampa, FL, USA}
\\[1pt]
\textit{$^{7}$Massachusetts General Hospital, 
Boston, MA, USA}
\\[1pt]
\textit{$^{8}$Department of Epidemiology and Population Health,
Albert Einstein College of Medicine,
Bronx, NY, USA}
\\[1pt]
\textit{$^{9}$Department of Environmental Medicine, 
Icahn School of Medicine at Mount Sinai,
New York, NY, USA}
\\[1pt]
\textit{$^{10}$Department of Gynecology and Obstetrics,
Emory University School of Medicine,
Atlanta, GA, USA}\\[1pt]
\textit{$^{11}$Department of Biostatistics,
University of Michigan School of Public Health,
Ann Arbor, MI, USA}\\[1pt]
\textit{$^{12}$Kaiser Permanente Division of Research,
Oakland, CA, USA}\\[1pt]
\textit{$^{13}$Department of Biostatistics,
Vanderbilt University Medical Center,
Nashville, TN, USA}\\[1pt]
\textit{$^{14}$Department of Biomedical Data Science,
Dartmouth College Geisel School of Medicine, 
Hanover, NH, USA}\\[1pt]
\textit{$^{15}$University of Colorado - Anschutz Medical Campus
Aurora, CO, USA}\\[1pt]
\textit{$^{16}$ Brown Center for the Study of Children at Risk
Providence, RI, USA}\\[1pt]
\textit{$^{17}$Department of Health Sciences,
Northeastern University Bouvé College of Health Sciences, 
Boston, MA, USA}\\[1pt]
\textit{$^{18}$Avera Research Institute
Sious Falls, SD, USA}\\[1pt]
\textit{$^{19}$Department of Pediatrics and Internal Medicine,
University of South Dakota Sanford School of Medicine,
Sioux Falls, SD, USA}\\[1pt]
}
% E-mail address for correspondence

% Running headers of paper:
\maketitle

% Add a footnote for the corresponding author if one has been
% identified in the author list
\footnotetext{To whom correspondence should be addressed.}

% Add a footnote for the corresponding author if one has been
% identified in the author list
\footnotetext{To whom correspondence should be addressed.}
\begin{abstract}
{Integrative analysis of datasets generated by multiple cohorts is a widely-used approach for increasing sample size, precision of population estimators, and generalizability of analysis results in epidemiological studies. However, often each individual cohort  dataset  does not have all variables of interest for an integrative analysis collected as a part of an original study.  Such cohort-level missingness poses methodological challenges to the integrative analysis since missing variables have traditionally: (1) been removed from the data for complete case analysis; or (2) been completed by missing data interpolation techniques using data with the same covariate distribution from other studies. In most integrative-analysis  studies, neither approach is optimal as it leads to either loosing the majority of study covariates or challenges in specifying the cohorts following the same distributions. We propose a novel approach to identify the studies with same distributions that could be used for completing the cohort-level missing information. Our methodology relies on (1) identifying sub-groups of cohorts with similar covariate distributions using cohort identity random forest prediction models followed by clustering; and then (2) applying a recursive pairwise distribution test for high dimensional data to these sub-groups. Extensive simulation studies show that cohorts with the same distribution are correctly grouped together in almost all simulation settings.  Our methods' application to two ECHO-wide Cohort Studies reveals that the cohorts grouped together reflect the similarities in study design. The methods are implemented in R software package \texttt{relate}.}
{Missing data; Meta analysis; Data integration; Random forest;
Clustering; Distribution similarity.}
\end{abstract}

\section{Introduction}
\label{sec1}

%\textcolor{orange}{BL: I am adding/editing text in orange. NOTE I am changing each of the individual level cohort language to ``independent contributing study (ICS)'' to be congruent with Lesko et al paper as the combination of studies does not necessarily have to be just cohort studies. However, I don't feel strongly } 

Collaborative study designs (CSD) in which individual data from multiple independent contributing  studies (ICS) have become more common (\citealp{lesko2018collaborative}). The advantages of combining data from multiple sources include increased statistical power through larger sample sizes, increased potential for investigating effect heterogeneity, and increased value by capitalizing on resources already used by the existing studies. However, despite numerous examples of CSDs (Table 1 in \citeauthor{lesko2018collaborative}, \citeyear{lesko2018collaborative}) the issue of missing data interpolation at the study level has not been a focus of much research.

The ECHO-wide Cohort Study harmonizes extant data from 69 ongoing ICS across the U.S. to study environmental factors associated with child health. Each of these ICS were initiated with heterogeneity in the research design due to the each of their overall research goals. Thus, the  extant data that has been collected may have been done with different ``tools'' (i.e., questionnaires, assays, etc.) and may or may not have overlapping data elements with the other studies being pooled to form the ECHO-wide Cohort. While the collected data have now been harmonized, the analysis of these extant data still creates an issue of how to deal with missingness at the ICS level. For example, if we have a causal question about an exposure, $X$ on $Y$, we need data on all the potential confounders to have the assumption of exchangeability hold. Yet, if among the 69 cohorts, only a subset have the full  set of confounding variables, it is unclear how to handle this missingness. The potential options include (but not limited to): (1) do a complete case analysis and drop the ICS that have not collected the full set of potential confounders; (2) pool all the data, conduct imputation, and ignore the issue that each of the ICS likely represent different populations and therefore the joint distributions may be heterogeneous between cohorts; or, (3) impute by grouping ICS into subsets based upon some meta-level information such as who, how, and when individuals were recruited into each of the studies. This last part is difficult as it is not always clear how the samples are different. Additionally, as multiple imputation relies on the joint distribution of the variables pertinent to the analyses, one may erroneously group ICS together when they should not be and vice versa (\cite{azur2011multiple}).

The question of cohort-level missing data inference in CSD has been discussed by \citeauthor{kundu2019} (\citeyear{kundu2019})  in the context of regression modeling of an outcome given a set of covariates. However, these methods assume the similarity in distribution of covariates across all studies, which, in practice, often may not hold. We consider an alternative approach in which imputation is based on clusters that suggest that the joint distribution of the observed data is not different from one another. Therefore, we propose an approach that clusters the ICS based upon the joint distribution of the pertinent variables that are observed. This is analogous to assuming that the  data are missing at random (MAR) given the joint distribution of observed pertinent variables (\citealp{rubin1976inference, little2019statistical}). The approach we consider is a two-stage approach, where we first identify groups by the joint distribution of observed variables, and then utilize imputation to account for the ICS level missing covariates. This approach is computationally feasible over directly testing all pairwise ICS comparisons which would be a brute force approach that in large collaborative studies becomes increasingly computationally expensive especially when each of the ICS has increasingly larger sample size. %and should one rely on a resampling or permutation approach for the variability rather than asymptotic. 

\section{Methods}
\label{sec2}

%\textcolor{orange}{NOTE: some of the language that is used has been collaborative study design, pooled individual data analysis, individual pooled data meta analysis. My weak preference is for collaborative study design and pooled individual data. I generally shy away from  ``meta'' analysis as my feeling it is more of using data that has already been published on and in this specific case multiple studies that have published on the scientific question but then pooling the individual data for a meta analysis rather than a meta analysis on the individual study results.}

\begin{figure}[!p]%to change at the end
\includegraphics[width=1\textwidth, height = 0.3\textheight]{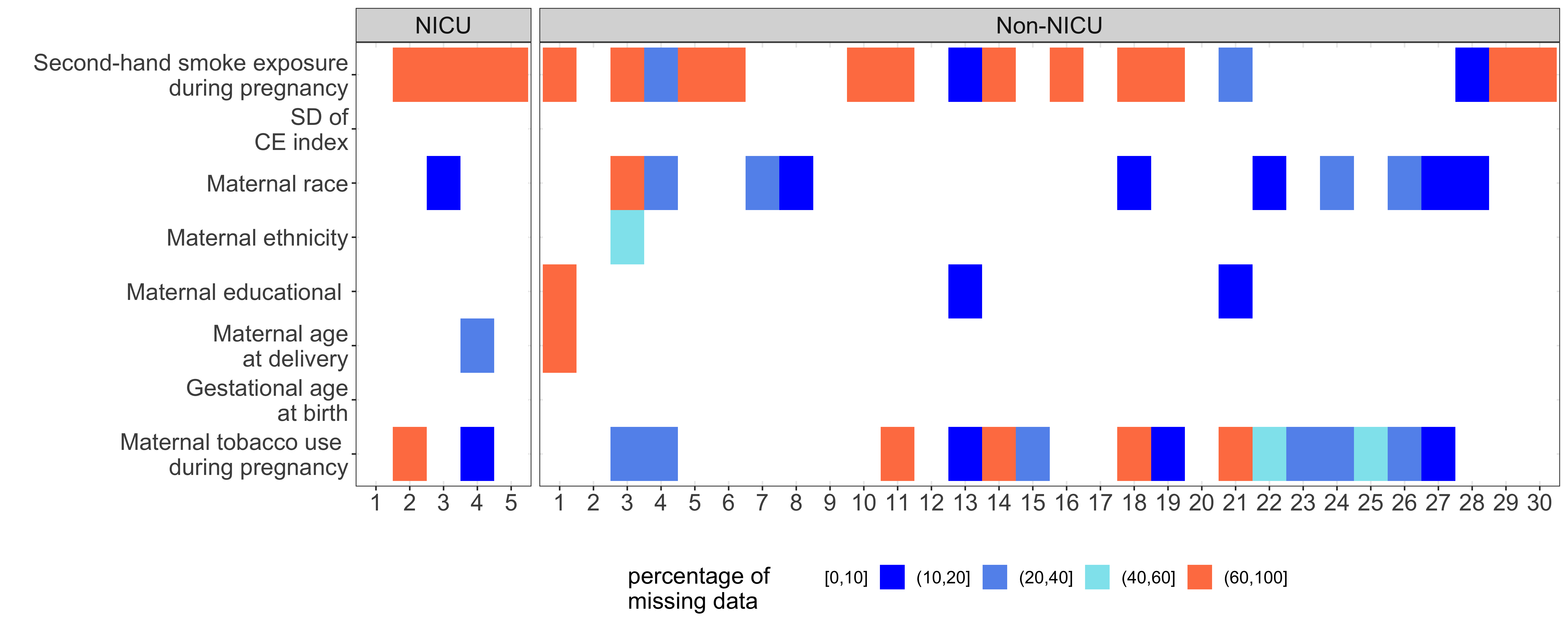}
\caption{{\small  Missing data pattern for the Neighborhood-level exposures to environmental and social stressors on pre-term birth study \cite{martenies2022associations}. The color scale represents the proportion of missing covariates (y-axis) for each cohort (x-axis). Cohorts are grouped by corresponding study design: (1) NICU - children who stayed at neonatal intensive care unit; (2) non-NICU - all other children.}}
\label{fig:Miss_dataEC0206}
\end{figure}

We introduce a novel approach for identifying  sub-groups of cohorts with the same joint distribution on  the measured covariates for use in pooled individual data analyses to allow ICS-level missing data interpolation. We start by introducing some notation. We assume there are $K$ ICS, which for simplicity we assume are all of a cohort study design. We denote individuals within each cohort $i=1, \dots, I_k, k = 1, \dots, K$. Suppose that for each individual we observe a set of variables  $\boldsymbol z_{ik} = \{z_{pik}\}_{p=1}^P$ of length $P$, which combines all measures of interest for a CSD analysis and may include an outcome, exposures, confounders, and effect modifiers (see Figure \ref{fig:Miss_dataEC0206}). We further assume that the data for a sub-set of variables $J \subset \{1, 2, \dots, P\}$ %\textcolor{orange}{NOTE/Question: shouldn't this be $<P$ since if a cohort is missing all relevant variables then there is no distribution to compare to other cohorts?} 
is either not collected  by at least one study or has a large percentage of missing data, which would make cohort-level interpolation statistically inappropriate. 

We consider a two-stage approach  called \emph{REcursive muLtivariAte TEsting for cohort clustering (relate)}, where we first (1) identify sub-groups of cohorts based on their similarity in the variables $\boldsymbol z_{ik}$ of interest for the CSD analysis and then (2) formally test the equivalence in distributions for these sub-groups. Our sub-grouping approach substantially reduces the number of tests needed for the distributional equivalence and the corresponding computational time. The full procedure is summarized in Algorithm \ref{alg:Algorithm 1}.

\begin{algorithm}\label{alg:Algorithm 1}
\caption{Cohort subgroups identification.}\label{alg:three}
\KwData{Participant-level table with  covariates for meta-analysis $\boldsymbol z_{ik}$ and cohort membership ID ($k=1, \dots, K$); significance level $\alpha$}
\KwResult{Cohort groups with same distribution on pooled data covariates}
\textbf{Stage 1: Identifying sub-groups of cohorts with similar  distributions}\\
 \textbf{Step 1}: Random forest classification\;
 \hspace{5mm} (i) Run multi-class random forest model to predict cohort ID using pooled data analysis covariates as predictors\;
 \hspace{5mm} (ii) Calculate pairwise random forest distance between participant-level observations\;
 \hspace{5mm} (iii) Average participant-level pairwise random forest distance within each cohort\;
 \textbf{Step 2}: Clustering to identify sub-groups of similar cohorts\;
 \hspace{5mm} (i) Run hierarchical clustering algorithm on cohort-level random forest distance matrix\;
 \hspace{5mm} (ii) Extract clustering dendrogram (D) to identify cohorts that group together\;
 \textbf{Stage 2: Recursive pairwise distribution test for high dimensional data}\\
 Biswas and Ghosh (BG) procedure on clustering dendrogram to test cohort similarity.
\end{algorithm}

\subsection{First stage---identifying sub-groups of cohorts with similar  distributions}

We start measuring the cohort similarity using a multi-class random forest model. Specifically, given a  vector of $\boldsymbol z_{ik}$ measures over a total of $KI_k$ observations in $K$ studies, we build a random forest model to predict the cohort membership, $k=1, \dots, K$All covariates of interest for the pooled analysis are used as predictors in the random forest model, where cohort-level missing covariates are interpolated within the model using the random forest imputation (\texttt{na.action="na.impute"} option in \texttt{R} function \texttt{rfsrc}).
In this approach, the model decision tree, $\mathcal T_b$, is built for each bootstrap sample $\mathcal I_b \subset \{1, \dots, KI_k\}$ of observations, where $~b = 1, \dots, B$.

To model the similarity between the cohorts based on their covariate distribution, we propose to use the random forest proximity measures. Random forest proximity can then be calculated using the measures of closeness between two observations on the decision tree $\mathcal T_b$ and averaged across $B$ trees (\citealp{breiman2003manual}).
%0-far 1 - close
The traditional random forest proximity measures the frequency at which two observations  end up in the same terminal node of the decision tree $\mathcal T_b$. However, this proximity does not account for the tree hierarchical structure and two observations from different nodes are treated equivalently regardless of the split location (\citealp{mantero2021unsupervised}). To overcome this drawback, we utilize the random forest dissimilarity introduced by  \citet{mantero2021unsupervised} that counts the number of splits  on the tree path between a pair of observations   $\boldsymbol z_{ik}$ and $\boldsymbol z_{jl}$ terminal nodes. Formally, let $S(\boldsymbol z_{ik}, \boldsymbol z_{jl}, \mathcal T_b)$ be the the minimum number of splits on the path from the terminal node containing $\boldsymbol z_{ik}$ to the terminal node containing $\boldsymbol z_{jl}$ in $\mathcal T_b$ such that the path includes at least one common ancestor node of $\boldsymbol z_{ik}$ and $\boldsymbol z_{jl}$. Let $R(\boldsymbol z_{ik}, \boldsymbol z_{jl}, \mathcal T_b)$ be the the minimum number of splits on the path from the terminal node containing $\boldsymbol z_{ik}$ to the terminal node containing $\boldsymbol z_{jl}$ in $\mathcal T_b$ such that the path includes the root node. Then the random forest dissimilarity measure between  $\boldsymbol z_{ik}$ and $\boldsymbol z_{jl}$ is given by 
\begin{equation}\label{eq:distance_measure}
d_{ik, jl} = D(\boldsymbol z_{ik}, \boldsymbol z_{jl})  = \frac{1}{B}\sum_{b=1}^B \frac{S(\boldsymbol z_{ik}, \boldsymbol z_{jl}, \mathcal T_b)}{R(\boldsymbol z_{ik}, \boldsymbol z_{jk}, \mathcal T_b)}.   
\end{equation}

To identify the sub-groups of cohorts with similar distributions, we first calculate the distance matrix with pair-wise random forest dissimilarity (\ref{eq:distance_measure})  between observations $D = \{d\}_{ik, jl}$, where  $i=1, \dots, I_k$ are the individuals within cohort $k$ and $j=1, \dots, I_l$ are the individuals within cohort $l$; $k, l = 1, \dots, K$. We then average these distances across all observations within the same cohort to obtain pair-wise random forest dissimilarity between cohorts, $\bar D = \{\bar d\}_{k, l}$, such that 
\begin{equation}\label{eq:distance_measure_cohorts}
\bar d_{k, l} = D(\boldsymbol z_{k}, \boldsymbol z_{l}) = \frac{1}{I_k}\frac{1}{I_l}\sum_{i=1}^{I_k}\sum_{j=1}^{I_l} D(\boldsymbol z_{ik}, \boldsymbol z_{jl}).
\end{equation}
The cohort-level pairwise distance matrix is then used in hierarchical clustering algorithm to group the cohorts  (\citealp{legendre2012numerical, gordon1999classification}). This hierarchical dendrogram is used to guide the second stage and reduce the number of pairwise comparisons.

Our approach assumes that observations in two or more distinct cohorts that come from the same distribution will fall close on the random forest decision tree and thus will have smaller dissimilarity measure as opposed to the variables that come from the different distributions. Specifically, for all $i, j$ we expect $d_{ik, jl}$ to be small  if cohorts $k$ and $l$ have the same distribution and  large  if cohorts $k$ and $l$ come from different distributions. Averaging  random distances across all observations within the same cohort is necessary to characterize dissimilarities at the cohort level that is of interest for the current study. Finally, the clustering step allows to efficiently identify the similarity structure between the cohorts by constructing a hierarchical dendrogram.

\subsection{Second stage---recursive pairwise distribution test for high dimensional data}

The  first stage of the proposed algorithm identifies the hierarchical similarity structure among the cohorts with a goal of reducing the number of tests required to formally test  for the equivalence of multivariate distributions in the second stage. This is done by applying the  two-sample non-parametric test for the multi-variate distributional equivalence developed by Biswas and Ghosh (BG) (\citealp{biswas2014nonparametric}). We recursively apply the BG test according to the clustering dendrogram. This non-parametric test is designed for testing the equivalence of two distributions $F$ and $G$ (not necessarily Gaussian) by comparing the average Euclidean distances  between the samples from these distributions. Formally, consider the independent  random vectors $\boldsymbol {X}, \boldsymbol {X}_* \overset{iid}{\sim} F$ and $\boldsymbol {Y}, \boldsymbol {Y}_* \overset{iid}{\sim} G$.  Let $D_{FF}$, $D_{GG}$ and $D_{FG}$ be the distributions of Euclidean distances between random vectors $\|\boldsymbol {X} - \boldsymbol {X}_*\|$, $\|\boldsymbol {Y} - \boldsymbol {Y}_*\|$, $\|\boldsymbol {X} - \boldsymbol {Y}\|$, respectively, and $\mu_{FF}$, $\mu_{GG}$, $\mu_{FG}$ their means. Let bivariate vectors  $(\|\boldsymbol {X} - \boldsymbol {X}_*\|, \|\boldsymbol {X} - \boldsymbol {Y}\|)$ and $(\|\boldsymbol {Y} - \boldsymbol {Y}_*\|, \|\boldsymbol {X} - \boldsymbol {Y}\|)$ follow distributions $D_F$ and $D_G$, respectively, and $\mu_{D_F}$, $\mu_{D_G}$ be their respective means. Then if $F$ and $G$ differ, so do $D_F$ and $D_G$, and visa versa.   Therefore, instead of testing $H_0: F = G$, the equivalent null hypothesis $H_0: \mu_{D_F} = \mu_{D_G}$ is tested. The corresponding test statistic is calculated from the samples $\boldsymbol{x}_1, \dots, \boldsymbol{x}_m \overset{iid}{\sim} F$ and $\boldsymbol{y}_1, \dots, \boldsymbol{y}_n \overset{iid}{\sim} G$ as
\begin{equation}\label{eq:BG_test}
\begin{aligned}
    \hat{\mu}_{D_F} = \Big[\hat{\mu}_{FF} = {m\choose 2}^{-1} \sum_{i=1}^m \sum_{j=i+1}^m \|\boldsymbol x_i -  \boldsymbol x_j\|; \hat{\mu}_{FG} = (mn)^{-1} \sum_{i=1}^m\sum_{j=1}^n \|\boldsymbol x_i - \boldsymbol y_j\| \Big] \\
    \hat{\mu}_{D_G} = \Big[\hat{\mu}_{FG} = (mn)^{-1} \sum_{i=1}^m\sum_{j=1}^n \|\boldsymbol x_i - \boldsymbol y_j\|; \hat{\mu}_{GG} = {n\choose 2}^{-1} \sum_{i=1}^n \sum_{j=i+1}^n \|\boldsymbol y_i -  \boldsymbol y_j \| \Big].
\end{aligned}
\end{equation}

\begin{figure}[!p]%to change at the end
\includegraphics[width=1\textwidth, height = 0.3\textheight]{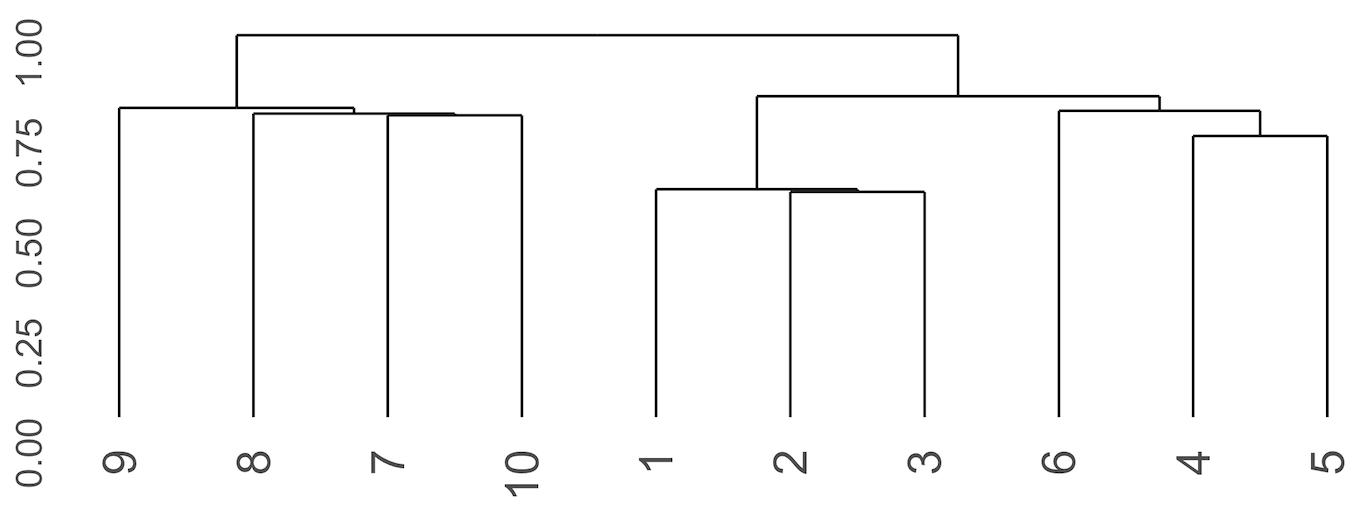}
\caption{{\small  Example clustering dendrogram for the cohort similarity testing. The numerical labels on the dendrogram terminal nodes correspond to the individual cohorts identification numbers. }}
\label{fig:BGTestExample}
\end{figure}

An example of this procedure for the case of testing distributional similarity among $K=10$ simulated cohorts is schematically represented in Figure \ref{fig:BGTestExample}. Let the data from cohorts $1, 2, \dots,  10$ follow distributions $F_1, F_2, \dots, F_{10}$, respectively. Note that each of the distributions within the following sets are from the same underlying distribution: \{$F_1, F_2, F_3$\}, $\{F_4, F_5, F_6\}$, and $\{F_7, F_8, F_9, F_{10}\}$. We start by applying the BG test (\ref{eq:BG_test}) to the data at the same height on the terminal nodes of the dendrogam. Specifically, the following null hypotheses are tested:
\begin{equation}
    H_{0}^1: F_2 = F_3;~ H_{0}^2: F_4 = F_6;~ H_{0}^3: F_9 = F_{10};~ H_{0}^4: F_7 = F_{9};~ H_{0}^5: F_7 = F_{10}.
\end{equation}

If  null hypotheses $H_{0}^1, \dots, H_{0}^5$ are not rejected at the pre-specified $\alpha$-level, then corresponding cohorts (1) $2$ and $3$; (2) $4$ and $6$; (3) $9$ and $10$; (4) $7$ and $9$; (5) $7$ and $10$ have same distributions and are combined in further analysis. If either $H_0$ were rejected, then corresponding cohorts have different distributions and no further testing is performed. For the combined data, the algorithm then moves to the next height of the dendrogram and the following hypotheses are tested
\begin{equation}
    H_{0}^{6}: F_{2,3} = F_1;~ H_{0}^{7}: F_{4,6} = F_5;~ H_{0}^8: F_{7,9,10} = F_8,
\end{equation}
where $F_{2,3}$, $F_{4,6}$ and $F_{7,9,10}$ are the distributions of the combined cohorts (a) $2$ and $3$, (b) $4$ and $6$, (c) $7$, $9$, and $10$, respectively.  The procedure is repeated until  the cohorts'  similarity is tested at the higher heights of the dendrogram, and so on until either the test rejects the null hypothesis and we stop grouping cohorts or all cohorts are combined. 

For each comparison using the BG test, the distribution being tested is on the overlapping observed variables. Therefore, each comparison at the first level of the dendrogram may be on different subsets of the $P$ variables. However, once two cohorts have been combined there are two different options for comparing a set of combined cohorts against a cohort at a higher level of the dendrogram (e.g., $H_{0}^8: F_{7,9,10} = F_8$). The first option is to only keep the comparison in the BG test to the overlapping variables and non-missing data even if for instance a missing variable in cohort 9 could be imputed due to combining with cohorts 7 and 10. The second approach is to impute the cohort level missing variable in cohort 9 by utilizing the information of the joint distributions of cohorts 7 and 10 with the non-missing variables in cohort 9. Provided that the assumption of the procedure is: given the observed variables, we cannot reject the null hypotheses of $H_{0}^3$, $H_{0}^4$, and $H_{0}^{5}$. We evaluate the accuracy of both approaches in Simulation section.

%we therefore embrace these results and suggest the second approach of imputing missingness in cohort 9.

Additionally, in following our approach (see Algorithm  S1 in the Supplementary Materials), there is the possibility that a cohort could be considered to be able to fit into two different clusters. Using the prior example, let us assume that the null hypotheses $H_{0}^3$ and $H_{0}^5$ were not rejected but $H_{0}^4$ was rejected at the $\alpha$-level. This would suggest that cohort 10 could be combined with either cohort 7 or with cohort 9, but cohort 7 and 9 are different enough from one another by the BG test and thus forming 2 clusters at this time point. The decision at this point is how to combine cohort 10 with either cohort 7 or cohort 9. We prioritize combining cohorts in this situation by the BG test statistic or equivalently by the magnitude of the p-value. The higher the p-value the closer the two cohorts are to one another. Therefore if the p-value for $H_{0}^3$ was $0.89$ and for $H_{0}^5$ was $0.41$ then cohort 10 would be combined with cohort 9 under this prioritization rule.

%\begin{figure}[!p] %to change at the end

\section{Simulation study}

\subsection{Simulation settings}

To evaluate the ability of the proposed \texttt{relate} method to detect cohorts with similar distributions, we simulate 10 cohorts with 15 covariates in each. For each simulation, the sample size of each cohort is randomly selected from the set $\{30, 50, 100, 200, 500, 700, 750, 900, 1000, 1200\}$. The distribution of the covariates is as follows: 10 covariates follow a multivariate normal distribution, two covariates are binary (above versus below the median of the normal distribution), one covariate is binomial, and the last two covariates are multinomial. Using these settings, we tested the following five simulation cases with 100 simulation per each scenario:\\
%\begin{enumerate}

\textbf{Case 1: same distribution}. Cohorts 1-10 are generated from one set of distribution parametersfor each type of distribution. \\
\textbf{Case 2: two distributions}. Cohorts 1-8 and 9-10 are generated using two  sets of parameters for each type of distribution, where for the normally distributed  covariates the two parameter sets have different means but the same covariance. \\
\textbf{Case 3: three distributions}. Cohorts 1-8, 4-6, and  7-10 are generated using three sets of  parameters for each type of distribution, where for the normally distributed  covariates the three parameter sets have different means but the same covariance. \\
\textbf{Case 4: two distributions}. Cohorts 1-5 and 6-10 are generated using two sets of parameters for each type of distribution, where for the normally distributed  covariates the two parameter sets have different means but the same covariance. \\
\textbf{Case 5: four distributions}. Cohorts 1-3, cohorts 4 and 5, cohorts 6 and 7, and cohorts 8-10  are generated using four sets of parametersfor each type of distribution, where for the normally distributed  covariates the two parameter sets have different means and different covariance. \\
%\end{enumerate}    
The distribution parameters for each simulation case are summarized in  Table S1 in the Supplementary Materials.

To mimic the ECHO data, we add a layer of missing at random (MAR), where four cohorts are missing four covariates, with a percentage of 40\%,  and 60\% of data missing in these covariates, respectively. Next, we  randomly choose the number of cohort-level missing covariates  according to  the following  scenarios: (Scenario 1) 2 cohorts missing 2 covariates; (Scenario 2) 2 cohorts missing 4 covariates; (Scenario 3) 5 cohorts missing 2 covariates; (Scenario 4) 5 cohorts missing 4 covariates; (Scenario 5) 5 cohorts missing 6 covariates. These numbers are chosen to examine the low (scenario 1-2), medium (scenario 3), and high (scenario 4-5) levels of missingness. These scenarios are chosen to quantify the potential decrease of the proposed \texttt{relate} method  performance with increasing missingness level.

\subsection{Simulation results}

The performance of the \texttt{relate} method is evaluated based on the following metrics: (1) correct number of detected cohort clusters, which corresponds to the number of distributions in each simulation scenario; (2) the number of cohorts assigned to the wrong cluster of distributions, which counts whether the cohorts assigned to the same cluster originate from the same distribution; %(3) the accuracy of the test, defined as the average number of correctly identified distributionss; 
(3) the time (in seconds) to fit the random forest model and hierarchical clustering; (4) the time to conduct the BG test per simulation. all simulations were performed on a cluster with 128GB-512GB per node RAM (\url{https://bit.ly/VCUCluster}).

The results for all  simulation scenarios are summarized in Tables \ref{table:SimulationResults_noimpute} and \ref{table:SimulationResults} for the case without and with missing data imputation at each step of recursive BG test, respectively. When recursive pairwise distribution BG test is performed without interpolating missing data for the cohorts that are combined at each step of recursion, the overall accuracy of the \texttt{relate} method is substantially higher. Specifically, \texttt{relate} identifies clusters of same distributions with over $90\%$ accuracy for the first 4 simulation cases with  $1$ to $3$ different distribution clusters in $10$ simulated cohorts. As expected, the accuracy is overall lower when the missingness level increases. The number of correctly identified clusters is lower for simulation case $5$, where data from $4$ different distributions is generated. The performance of this method for this simulation case is especially low ($54\%$ accuracy) when the number of cohort-level missing covariates is high.   This low performance may be explained by the low number of cohorts with similar distribution in each cluster ($2$ to $3$). 

When missing data is imputed at each step of the recursive BG test algorithm,  \texttt{relate} method identifies the clusters of similar distributions with over $99\%$ accuracy if there are $2$ cohort-level missing variables and up to a four distinct distributions.   With the larger number of missing variables, the number of cohort clusters increase which leads to the lower number of correctly defined distribution clusters. It should be noted that the rate for the wrong number of assignments is low, which means that similar distributions are combined correctly but a large cluster of similar distributions may be split into two or more distinct clusters.  However, with four distinct distributions out of the 10 cohorts, some cohorts are combined into incorrect clusters in $23-27\%$ of all simulation cases except for the case with five cohorts missing two covariates.  

The average running time for the stage 1 (random forest and clustering method) of approach is approximately 50 seconds, and the average time for stage 2 (BG testing for similarity in distributions) is approximately 40 seconds.   Together, these results indicate that the proposed \texttt{relate} methodology accurately detects clusters of similar distributions even when  multiple cohorts are missing a large number of covariates.  The run time for the method is on average under 2 minutes, which makes it feasible to implement on large-scale data sets.   

\begin{table}[!p]
\caption{{\small  Performance summary of \texttt{relate} without imputation while running BG test by each cohort-level covariate missingness scenario and the true number of distribution clusters over 100 simulations.  Results are summarized by (1) correct number of detected cohort clusters; (2) the number of cohorts assigned to the wrong cluster of distributions; 
(3) the time (in seconds) to fit the random forest model and hierarchical clustering; (4) the time to conduct the BG test per simulation.}}
\label{table:SimulationResults_noimpute}
\begin{minipage}{0.5\textwidth}
    \small
    \centering
    \begin{tabular}{|c|cc|cccc|c|cc|}
    \hline
%\toprule
%    \thead{Scenario\\(Missingness\\ level)} & \thead{Number\\ of\\ Cohorts} & \thead{Number\\ of\\ Covariates} & \thead{Sim.\\ Case} & \multicolumn{2}{c}{\thead{Number\\ of\\ Clusters}}& \thead{Wrong\\ Assignment} & \thead{Accuracy\\ (\% of correct\\ assignment)} & \multicolumn{2}{c}{Time/sim} \\ \hline ~& ~& ~& ~& Correct & Wrong & ~ & ~ & RF & BG\\ 
%\cmidrule(lr){5-6}\cmidrule(lr){9-10}
 Scenario &  Number  &Number  &Sim &Number of   &Number of &Number of &\% of  &\multicolumn{2}{c}{Time/sim}\\ 
 ~ &  of  & of  & Case & Correct   & Wrong& Correct& Wrong  &RF&BG\\ 
 ~& Cohorts & Covariates & ~& Clusters  & Clusters& Assignments & Assignments& \multicolumn{2}{c}{}\\ \hline
       \multirow{5}{*}{1} & \multirow{5}{*}{2} &\multirow{5}{*}{2} & 1 & 100 & 0 & 0 & 100 & 49.07 & 38.97\\ 
       ~& ~& ~& 2 & 100 & 0 & 0 & 100 & 60.75 & 49.42 \\
       ~& ~& ~& 3 & 100 & 0 & 0 & 100 & 51.75 & 33.27 \\
       ~& ~& ~& 4 & 100 & 0 & 0 & 100 & 57.28 & 15.89 \\
       ~& ~& ~& 5 & 98 & 2 & 29 & 71 & 32.9 & 12.82\\ \hline
        \multirow{5}{*}{2} & \multirow{5}{*}{2} &\multirow{5}{*}{4}& 1 & 100 & 0 & 0 & 100 & 49.9 & 71.60 \\
        ~&~&~& 2 & 100 & 0 & 0 & 100 & 43.5 & 52.22\\
        ~& ~& ~& 3 & 100 & 0 & 0 & 100 & 38.7 & 11.84 \\
        ~&~&~& 4 & 100 & 0 & 0 & 100 &  41.14 & 15.43 \\
        ~&~&~& 5 & 87 & 13 & 46 & 54 &36.02 & 13.35 \\ \hline
       \multirow{5}{*}{3} & \multirow{5}{*}{5} &\multirow{5}{*}{2}& 1 & 100 & 0 & 0 & 100 & 49.97 & 71.60 \\ 
       ~&~&~& 2 & 100 & 0 & 0 & 100 & 43.54 & 52.22 \\
       ~&~&~& 3 & 100 & 0 & 0 & 100 & 38.71 & 11.84 \\
       ~&~&~& 4 & 100 & 0 & 0 & 100 & 41.24 & 27.72 \\
       ~&~&~& 5 & 98 & 2 & 21 & 79 & 42.24 & 16.66 \\ \hline 
       \multirow{5}{*}{4} & \multirow{5}{*}{5} &\multirow{5}{*}{4}& 1 & 100 & 0 & 0 & 100 & 51.70 & 83.01 \\ 
       ~&~&~& 2 & 100 & 0 & 0 & 100 & 49.93 & 136.83 \\
       ~&~&~& 3 & 98 & 2 & 6 & 94 & 43.17 & 18.68 \\
       ~&~&~& 4 & 100 & 0 & 0 & 100 & 45.36 & 47.32 \\
       ~&~&~& 5 & 98 & 2 & 21 & 79 & 42.44 & 16.66 \\ \hline
       \multirow{5}{*}{5} & \multirow{5}{*}{5} &\multirow{5}{*}{6}& 1 & 100 & 0 & 0 & 100 & 87.14 & 79.79 \\ 
       ~&~&~& 2 & 99 & 1 & 1 & 99 & 81.72 & 88.30 \\
       ~&~&~& 3 & 98 & 2 & 9 & 91 & 53.79 & 14.15 \\
       ~&~&~& 4 & 100 & 0 & 0 & 100 & 53.67 & 22.40 \\
       ~&~&~& 5 & 90 & 10 & 46 & 54 & 52.06 & 14.29 \\ \hline
    \end{tabular}
\end{minipage}    
\end{table}

\begin{table}[!p]
\caption{{\small  Performance summary of \texttt{relate} with imputation while running BG test by each cohort-level covariate missingness scenario and the true number of distribution clusters over 100 simulations.  Results are summarized by (1) correct number of detected cohort clusters; (2) the number of cohorts assigned to the wrong cluster of distributions; 
(3) the time (in seconds) to fit the random forest model and hierarchical clustering; (4) the time to conduct the BG test per simulation.}}
\label{table:SimulationResults}
   \begin{tabular}{|c|cc|ccc|cc|}
%    \toprule
%      \thead{Scenario\\(Missingness\\ level)} & \thead{Number of\\ Missing \\Cohorts} & \thead{Number of\\Missing\\ Covariates} & \thead{Simulation\\ Case} & \thead{Correct\\ Number of\\ clusters} & \thead{Number of\\Wrong\\ Clusters} & \multicolumn{2}{c}{Time/sim} \\
%      ~& ~& ~& ~ & ~ & ~ & RF & BG\\ 
%\cmidrule(lr){7-8}
\hline
Scenario  &Number of  &Number of  &Simulation &Number of &Number of &\multicolumn{2}{c}{Time/sim}\\ 
  (Missingness& Missing & Missing  & Case  & Correct & Wrong  &\multicolumn{2}{c}{|}\\ 
 level)&  Cohorts& Covariates & ~ & Clusters &Clusters &RF&BG\\ \hline
       \multirow{5}{*}{1} & \multirow{5}{*}{2} &\multirow{5}{*}{2} & 1 & 100 & 0 & 49.07 & 53.50\\ 
        ~& ~& ~& 2 & 100 & 0 & 60.75 & 63.54 \\
        ~& ~& ~& 3 & 98 & 1 & 51.77 & 24.53\\
        ~& ~& ~& 4 & 99 & 0 & 57.28 & 24.77 \\
        ~& ~& ~& 5 & 100 &  25 & 32.96 & 21.69\\ \hline
        \multirow{5}{*}{2} & \multirow{5}{*}{2} &\multirow{5}{*}{4}& 1 & 81  & 0 & 49.97 & 73.79 \\
        ~&~&~& 2 & 91 & 0 & 43.54 & 61.85\\
        ~& ~& ~& 3 & 68 & 2 & 38.71 & 19.00 \\
        ~&~&~& 4 & 77 & 0 &   41.14 & 22.31 \\
        ~&~&~& 5 & 84 & 27 & 36.02 & 20.55 \\ \hline
       \multirow{5}{*}{3} & \multirow{5}{*}{5} &\multirow{5}{*}{2}& 1 & 96 & 0 & 47.25 & 73.71 \\ 
       ~&~&~& 2 & 96 & 0 &  46.29 & 83.38 \\
       ~&~&~& 3 & 100 & 0 & 38.49 & 22.19 \\
       ~&~&~& 4 & 100 & 0 & 41.24 & 27.72 \\
       ~&~&~& 5 & 100 & 1 & 37.39 & 21.92 \\ \hline 
       \multirow{5}{*}{4} & \multirow{5}{*}{5} &\multirow{5}{*}{4}& 1 & 68 & 0  & 51.70 & 119.92 \\ 
       ~&~&~& 2 & 65 & 0 & 49.93 & 133.43 \\
       ~&~&~& 3 & 85 & 1 & 43.17 & 39.93 \\
       ~&~&~& 4 & 68 & 0 & 45.36 & 69.29 \\
       ~&~&~& 5 & 95 & 6 & 42.44 & 25.84 \\ \hline
       \multirow{5}{*}{5} & \multirow{5}{*}{5} &\multirow{5}{*}{6}& 1 & 68 & 0 & 87.14 & 55.28 \\ 
       ~&~&~& 2 & 40 & 0 & 81.72 & 43.70 \\
       ~&~&~& 3 & 64 & 1  & 53.79 & 22.44 \\
       ~&~&~& 4 & 64 & 1 & 53.67 & 22.40 \\
       ~&~&~& 5 & 91 & 23 & 52.06 & 14.29 \\ \hline
       
    \end{tabular}
\end{table}

\section{Applications}
\label{sec3}

We apply our method to the  previously collected (or extant data) from multiple cohorts participating in the Environmental Influences on Child Health Outcomes (ECHO) project (\citealp{gillman2018environmental}).   We consider two ECHO studies that examine: (1) the effect of neighborhood-level exposures to  environmental and social stressors on pre-term birth; and (2) the relationship of  cardiometabolic pregnancy complications to offspring autism spectrum disorder (ASD) traits. We define a covariate to be entirely missing if it was either (1) not collected by a cohort or (2) had $\geq 60\%$ missingness level. We compare three  cohort-level imputation methods: (1) pooled cohort imputation; (2) similar study design imputation; and (3) \texttt{relate} imputation. The cohort-level imputation methods are compared in terms of the differences on the regression coefficients estimation and inference. We assume the data are missing at random (MAR) and use the multiple imputation by chained equations method (10 imputations and each with 10 iterations) to impute the data (\citealp{donders2006gentle, stuart2009multiple}).

\subsection{Neighborhood-level exposures to  environmental and social stressors on pre-term birth}

The neighborhood-level exposures to  environmental and social stressors study uses the extant data from 35 cohorts. Study inclusion criteria have been described previously \cite{martenies2022associations}.  
%Individual study cohorts were eligible for this analysis when more than 30 participants had both residential history and perinatal outcome data. Participant addresses were geocoded in ArcGIS Pro Streetmap Premium Geocoder. Over 94\% of addresses had a high-quality match (point or specific street address) which was required for inclusion in this analysis. We then assigned a census tract identifier to each participant address using the 2010 census tract boundaries.  
The primary outcome of interest in this study was the gestational age at birth.
The primary exposure was a combined exposure (CE) index that characterized the exposure to several environmental hazards and social stressors, at the level of the  census tract (\citealp{cushing2015racial}). The standard deviation of the primary exposure index, $\mbox{CE}_{sd}$ was used in the analysis. A priori selected confounders, all assessed at the individual level, included maternal age at delivery (continuous), race (categories: White; Black; and other race), ethnicity (categories: non-Hispanic, Hispanic), education level (categories: less than high school; high school degree, GED or equivalent; some college degree and above), tobacco use during pregnancy (categories: never; and ever), and second-hand cigarette smoke exposure during pregnancy (categories: never; and ever). The number of missing covariates for each cohort is presented in Figure \ref{fig:Miss_dataEC0206}.

The proposed \texttt{relate} model identified six clusters of similar cohorts. One cluster contained only neonatal intensive care unit (NICU) cohorts, which are expected to be similar by the original study design. There were also two 
clusters comprised of only one cohort each: (1) a general population cohort of healthy women in Seattle, Washington (Global Alliance to Prevent Prematurity and Stillbirth, GAPPS); and (2) a Los Angeles, California-based cohort of pregnant women from low-income, urban minority populations (Maternal and Developmental Risks from Environmental and Social Stressors, MADRES). While GAPPS did not have any cohort-level missing covariates, maternal self-reported active and/or second-hand smoking during pregnancy were missing in MADRES. Therefore, MADRES cohort was not used in the analysis since we were unable to identify any cohort with similar distribution to perform imputation. For the imputation from the cohorts with similar study design, we imputed the data separately within the  five pre-term/NICU cohorts and within the remaining 30 cohorts. 

To predict gestational age at birth, we fitted the linear mixed models using \texttt{gee} function in \texttt{R} with cohort indicator as a random effect. The estimated regression coefficients and corresponding significance values for each predictive model are presented in Table \ref{table:EC206_GA}.  The values of estimated coefficients are similar across the three interpolation types. The significance level of the estimated regression coefficients is, however, dramatically different  for the pooled cohort imputation, where all cohorts are treated to have the same distribution. Specifically, the standard deviation of the primary exposure index, $\mbox{CE}_{sd}$, was significant for the pooled cohort imputation strategy and was not significant for NICU and \texttt{relate} imputation strategies. Similarly, the significance level of the estimated regression coefficients for the race category ``other" (i.e. not White or Black),  Hispanic origin, and second-hand tobacco smoking were significantly different for the total imputation strategy but not significant for the similar design and \texttt{relate} imputation strategies.   

\begin{table}[!]
\caption{{\small  Comparison of the estimated regression coefficients and significance values for prediction of  gestational age using (1) total: imputation from the cohorts with available data on a missing covariate; (2) similar design: imputation from the cohorts with similar study design; and (3) RF asymptotic: imputation from the clusters of similar cohorts identified by the proposed random forest method.}}
\centering
    \begin{tabular}{|l|l|l|l|l|l|l|}
    \hline
       ~& \multicolumn{2}{|c|}{Total}& \multicolumn{2}{|c|}{Similar design}& \multicolumn{2}{|c|}{RF asymptotic} \\ \hline
        ~ & Estimate & pvalue & Estimate & pvalue & Estimate & pvalue \\ \hline
        Intercept & 3268.733 & $<0.001$ & 3303.914 & $<0.001$ & 3278.078 & $<0.001$ \\ \hline
        $\mbox{CE}_{sd}$ & -49.450 & $<0.001$ & -53.456 & 0.076 & -52.976 & 0.076 \\ \hline
        Maternal age at delivery & -0.141 & 0.931 & -1.898 & 0.583 & -0.914 & 0.777 \\ \hline
        Maternal race &      &   &   &   &   &  \\ 
        Black & -279.605 & $<0.001$ & 242.950 & 0.001 & -257.456 & $<0.001$ \\ 
        Other & -82.213 & $<0.001$ & -84.643 & 0.308 & -83.757 & 0.309 \\ \hline
        Maternal ethnicity: Hispanic & 106.322 & $<0.001$ & 106.374 & 0.078 & 98.565 & 0.104 \\ \hline
        Maternal education &      &   &   &   &   &  \\ 
        High School & -11.735 & 0.768 & -24.709 & 0.527 & -16.639 & 0.653 \\ 
        College & 88.070 & 0.016 & 92.819 & 0.052 & 88.752 & 0.066 \\ \hline
        Maternal tobacco use &      &   &   &   &   &  \\ 
        during pregnancy & 33.685 & 0.312 & 150.790 & 0.336 & 153.632 & 0.342 \\ \hline
        Second-hand smoke &      &   &   &   &   &  \\ 
        exposure during pregnancy & -213.806 & $<0.001$ & -338.338 & 0.217 & -332.136 & 0.218 \\ \hline
    \end{tabular}
\label{table:EC206_GA}
\end{table}

\subsection{Cardiometabolic pregnancy complications to offspring autism spectrum disorder}

\begin{figure}[!p]%to change at the end
\includegraphics[width=1\textwidth, height = 0.3\textheight]{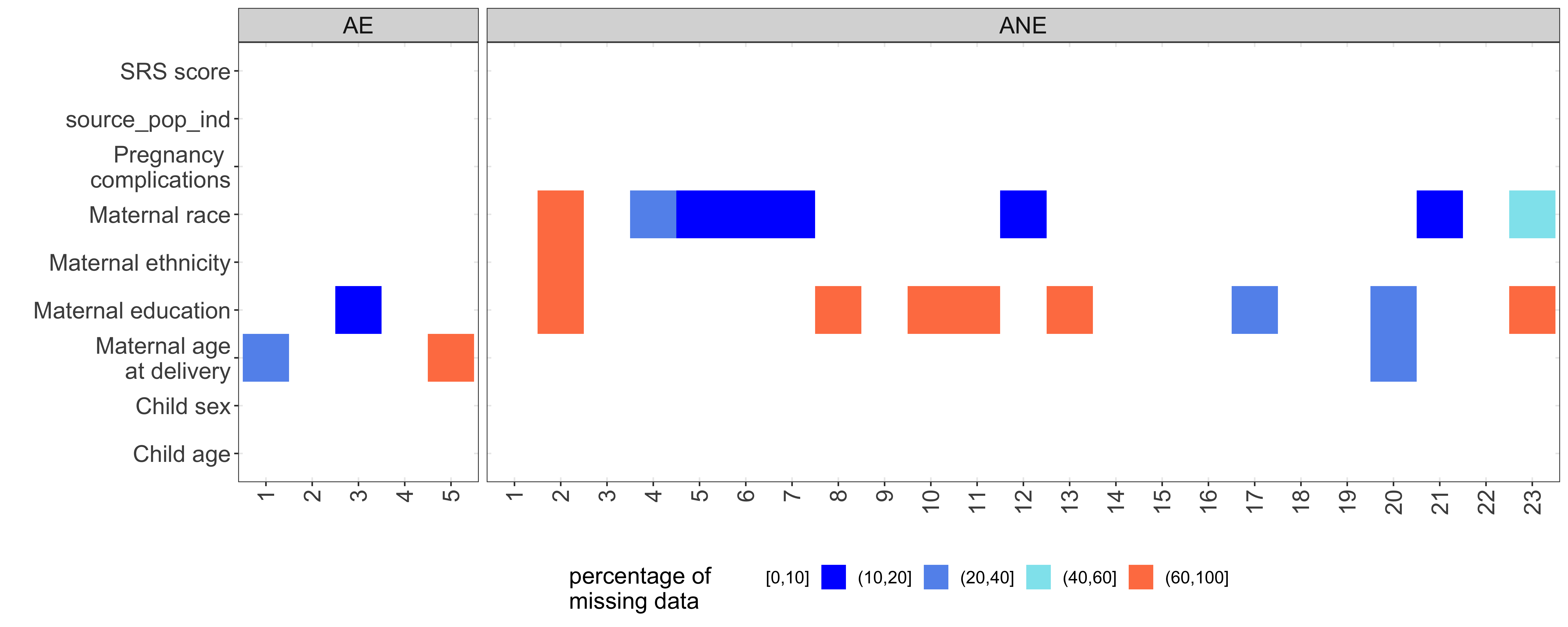}
\caption{{\small  Missing data pattern for the cardiometabolic pregnancy complications to offspring autism spectrum disorder study. The color scale represents the proportion of missing covariates (y-axis) for each cohort (x-axis). Cohorts are grouped by corresponding study design: (1) autism enriched (AE) - cohorts with higher proportions of children with autism disorder; (2) autism not enriched (ANE) - cohorts with a regular proportion of children with autism disorder.}}
\label{fig:Miss_dataEC0256}
\end{figure}

The cardiometabolic pregnancy complication for the offspring autism spectrum disorder study leveraged the data from 40 eligible ECHO cohorts. Cohorts inclusion and exclusion criteria are described in \cite{lyall2022cardiometabolic}. For the purpose of this analysis, 28 cohorts with $\geq 25$ observations per cohort were selected. The primary outcome of interest for this study was the Social Responsiveness Scale (SRS) score, a widely-used, validated measure of ASD-related traits (\citealp{constantino2012social}). Covariates of interest included maternal age (selected based on previously reported associations with ASD), any cardiometabolic pregnancy complication (pre-pregnancy obesity, gestational diabetes, hypertension during pregnancy, or preeclamsia), maternal race/ethnicity and educational level (as proxies for the  unmeasured sociodemographic factors such as access to healthcare and interpretation of social communication questions), child age at SRS  administration, and child sex (male/female). Missing data pattern for this cohort is presented in Figure \ref{fig:Miss_dataEC0256}.

Linear models (\texttt{R} function \texttt{lm}) were used to fit the models. The proposed \texttt{relate} approach identified seven groups of similar cohorts, two of which was comprised of only a single cohort: (1) Columbia Center for Children's Environmental Health (CCCEH); and (2) the study effects of potentially modifiable exposures during pregnancy and after birth (Project Viva). It is not surprising that the  CCCEH cohort was not grouped with other cohorts in the study: this cohort was unique comprised of women  identified as Dominican or African American and lived in Manhattan or South Bronx, NY. Both studies had cohort-level missing data and were removed from the analysis. For the imputation from cohorts with similar study design, we imputed the data within five autism-enriched cohorts and within the remaining  cohorts separately. %\textcolor{blue}{Not all cohorts with autism enriched/not enriched are in one cluster.}

The estimated regression coefficients and corresponding significance values are presented in Table \ref{table:EC256}. The estimated regression coefficients and corresponding significance levels are similar for the pooled cohorts and similar study design imputation methods, and are different from the proposed \texttt{relate} approach for the covariates: college education and race (American Indian and more than one race).  These results suggest that similarity in  study design does not necessarily imply similarity in covariate distribution for these cohorts. Therefore, imputing data from either pooled cohorts or large clusters of similar study design may often lead to violation of an assumption of similarity in data distributions used for imputation. While the underlying truth is unknown, we argue that using smaller clusters of cohorts with potentially similar distributions is a more reasonable approach to avoid changing the associations between exposures and outcomes.  

\begin{table}[!]
\caption{{\small  Comparison of the estimated regression coefficients for prediction of  SRS scores using (1) total: imputation from the cohorts with available data on a missing covariate; (2) similar design: imputation from the cohorts with similar study design; and (3) RF asymptotic: imputation from the clusters of similar cohorts identified by the proposed random forest method.}}
\centering
    \begin{tabular}{|l|l|l|l|l|l|l|}
    \hline
       ~& \multicolumn{2}{|c|}{Total}& \multicolumn{2}{|c|}{Similar design}& \multicolumn{2}{|c|}{RF asymptotic} \\ \hline
        ~ & Estimate & P-value & Estimate & P-value & Estimate & P-value \\ \hline
        Intercept & 40.361 & $<0.001$ & 42.252 & $<0.001$ & 40.474 & $<0.001$  \\ \hline
        Child's sex & -5.992 & $<0.001$ & -5.971 & $<0.001$ & -6.406 & $<0.001$ \\ \hline
        Child age & 0.723 & $<0.001$ & 0.747 & $<0.001$ & 0.848 & $<0.001$ \\ \hline
        Maternal race &   & & &  &  &  \\ \hline
        Black & 6.621 & $<0.001$ & 6.954 & $<0.001$ & 4.579 & $<0.001$ \\ \hline
        Asian & 6.589 & $<0.001$ & 7.006 & $<0.001$ & 5.256 & 0.001 \\ \hline
        Native Hawaiian or other Pacific & 0.544 & 0.926 & 2.899 & 0.681 & 1.359 & 0.852 \\ \hline
        American Indian or Alaska Native & 7.427 & 0.005 & 6.884 & 0.017 & 7.718 & 0.852 \\ \hline
        More than one race & 4.470 & 0.003 & 4.158 & 0.008 & 2.327 & 0.171 \\ \hline
        Other & 0.628 & 0.781 & 1.240 & 0.620 & -2.999 & 0.254 \\ \hline
        Hispanic & 2.436 & 0.022 & 2.788 & 0.008 & 2.499 & 0.041 \\ \hline
        Maternal age at delivery & 0.082 & 0.124 & 0.022 & 0.683 & 0.037 & 0.560 \\ \hline
        Maternal Education &      &   &   &   &   &  \\ 
        below high school & -2.200 & 0.148 & -2.015 & 0.195 & 0.709 & 0.654 \\ 
        some college & -4.975 & $<0.001$ & -5.131 & $<0.001$ & -1.463 & 0.326 \\ 
        Bachelors  & -10.882 & $<0.001$ & -10.612 & $<0.001$ & -7.803 & $<0.001$ \\ 
        Masters and above & -11.939 & $<0.001$ & -12.033 & $<0.001$ & -9.743 & $<0.001$ \\ \hline
        Pregnancy complication & 4.467 & $<0.001$ & 4.428 & $<0.001$ & 3.665 & $<0.001$ \\ \hline
    \end{tabular}
\label{table:EC256}
\end{table}

\section{Discussion}
\label{sec4}

We introduced a procedure for extending the methods for  missing data interpolation to the case of heterogeneous multi-cohort data where assumptions of  the joint distribution of the variables pertinent to the analyses may be violated.  The proposed approach  groups  ICS together, and then recursively assesses the similarity in  distributions between the pairs of grouped data. We implement our procedure in \texttt{R} software package \texttt{relate} and illustrate applications on two ECHO studies. Our extensive simulations suggest that the proposed methodology correctly groups similar cohorts when the number of cohort groups with similar distributions is small. 

The methodological novelty of our work is in the development of a combined approach for identifying groups with similar distributions using the random forest algorithm to define cohorts similarity based on the ability of covariates to predict cohort membership. The hierarchical clustering applied to the random forest distance further identifies the groups of cohorts with potentially similar distributions, which is formally tested using the BG test.  This approach  substantially reduces computational time  needed for pairwise group comparisons of distributional similarity and makes the methodology feasible to implement in pooled cohort studies with over 1,000 observations per cohort. Crucially, this is the first study that attempts to systematically interpolate large heterogeneous data with cohort-level missing covariate data. Our results show that the cohorts grouped using the proposed methodology  reflect the similarities in study design (i.e., grouping of all NICU cohorts together). 

Our method has several limitations. First, we assume that the data is missing at random (MAR); that is covariates were not collected because of resources and none of participating cohorts purposefully decided not to collect missing covariates. Second, while we used random forest approach followed by hierarchical clustering algorithm, our method's performance may change with alternative superlearning (i.e., a diverse library of data-adaptive algorithms to create the best weighted combination of candidate algorithms to further improve performance) algorithms (\citealp{naimi2018stacked}) and clustering approaches; selection of an optimal learning approach is a topic for future research. Third, as illustrated in two applied ECHO data examples, \texttt{relate} does not necessarily combine all cohorts together which results in singleton cohorts with missing data that have to be excluded from the analysis. On the other hand, this may also be interpreted as an advantage of the proposed method that identifies dissimilar cohorts that should not be grouped with other data for interpolation purposes.  Forth, we assume that the sample size is large enough to apply asymptotic distribution of the BG test statistic. 
Fifth,  based on our simulations, we suggest not to impute while running BG test and not no apply \texttt{relate} in small studies with 3-4 cohorts. However, obtaining theoretical foundations to support these findings is a topic for future research. Finally, we suggest that the relate cohort groupings are evaluated by the by the subject expert prior to imputation and final analysis.

%the accuracy of grouping assignments decreases with data heterogeneity and a larger number of cohort-level missing covariates. Second, the random forest algorithm may group the cohorts  

%(2) unsupervised learning (in clustering) and RF  -- other better algorithms to lear with many algorithms -- downside is without truth is real data  there is no guarantee it would be better that RF -- superlerner choose from multiple learning algorithms

%(3) missing at random assumption -- assume cohorts ddid not not purposefully decide not to collect information related to outcome -- reason did not collect is because of resoutces 

%(4) assume that we have sample sized large enpouth that asymptotics appply.

%(5) we drop some of the cohorts that are singleton clusters and loose data -- but an advantage is we do not make an assumption that these cohorts go with someone else

%%%%%%%%%%%%%%%%%%%%%%%%%%%%
%add to simulation section how much we loose when have many signleton clusters in simulations

\section{Software}
\label{sec5}

Software in the form of R code, together with a sample
input data set and complete documentation is available on
github (\url{https://yqzhong7.github.io/relate/}).

\section{Supplementary Material}
\label{sec6}

Supplementary material is available online at
% \href{http://biostatistics.oxfordjournals.org}%
% {http://biostatistics.oxfordjournals.org}.
\url{http://biostatistics.oxfordjournals.org}.

\section*{Acknowledgments}

% This study makes use of data generated by
%the ECHO. A full list of the
%investigators who contributed to the generation of the data is
%available from %\href{www.wtccc.org.uk}{www.wtccc.org.uk}.
%\url{www.wtccc.org.uk}.
The authors wish to thank our ECHO colleagues; the medical, nursing, and program staff; and the children and families participating in the ECHO cohorts. We also acknowledge the contribution of the following ECHO program collaborators:
ECHO Components—Coordinating Center: Duke Clinical Research Institute, Durham, North Carolina: Smith PB, Newby KL; Data Analysis Center: Johns Hopkins University Bloomberg School of Public Health, Baltimore, Maryland: Jacobson LP; Research Triangle Institute, Durham, North Carolina: Parker CB; Person-Reported Outcomes Core: Northwestern University, Evanston, Illinois: Gershon R, Cella D.
ECHO Awardees and Cohorts— Albert Einstein College of Medicine, Bronx, New York: Aschner J; Cincinnati Children's Hospital Medical Center, Cincinnati, Ohio:  Merhar SIndiana University, Riley Hospital for Children, Indianapolis, IN: Ren C; University of Buffalo, Jacobson School of Medicine and Biomedical Sciences, Buffalo, NY: Reynolds A; University of California, San Francisco: Keller R; University of Rochester Medical Center, Rochester, NY: Pryhuber G; University of Texas Health Sciences Center, Houston, TX: Duncan A; Vanderbilt Children’s Hospital, Nashville, TN: Moore P o	Children’s Hospital and Clinic Minneapolis, MN: Lampland A; Florida Hospital for Children, Orlando, FL: Wadhawan R; Medical University of South Carolina, Charleston, SC: Wagner C; 	University of Arkansas for Medical Science: Keller R; University of Florida College of Medicine, Jacksonville, FL: Hudak M; University of Washington, Seattle, WA : Mayock D; Wake Forest University School of Medicine, Winston Salem, NC: Walshburn L; University of Colorado Denver, Denver, CO: Dabelea D; Memorial Hospital of Rhode Island, Providence RI: Deoni S; New York State Psychiatric Institute, New York, NY: Duarte C; University of Puerto Rico, San Jaun, PR: Canino G; Avera Health Rapid City, Rapid City, SD: Elliott A; Kaiser Permanente Northern California Division of Research, Oakland, CA: Croen L; Boston Medical Center, Boston MA: Bacharier L; O’Connor G; Children’s Hospital of New York: New York, NY: Bacharier L; Kattan M; Johns Hopkins University, School of Medicine, Baltimore, MD: Wood R; Bacharier L; Washington University in St Louis, St Louis, MO: Rivera-Spoljaric K;	Henry Ford Health System, Detroit, MI: Johnson C; University of California Davis Mind Institute, Sacramento, CA: Hertz-Picciotto I; University of Pittsburgh, Pittsburgh, PA: Hipwell A;  Geisel School of Medicine at Dartmouth, Lebanon, NH: Karagas M; University of Washington, Department of Environmental and Occupational Health Sciences, Seattle, WA: Karr C; University of Tennessee Health Science Center, Memphis, TN: Mason A; Seattle Children’s Research Institute, Seattle, WA:  Sathyanarayana S; Children’s Mercy, Kansas City, MO: Carter B; Emory University, Atlanta, GA:Dunlop AL; Helen DeVos Children’s Hospital, Grand Rapids, MI: Pastyrnak S; Kapiolani Medical Center for Women and Children, Providence, RI: Neal C; 	Los Angeles Biomedical Research Institute at Harbour-UCLA Medical Center, Los Angeles CA: Smith L; 	Wake Forest University School of Medicine, Winston Salem, NC: Helderman J; Prevention Science Institute, University of Oregon, Eugene, OR: Leve L; George Washington University, Washington, DC: Ganiban J; Pennsylvania State University, University Park, PA: Neiderhiser J; Brigham and Women's Hospital, Boston, MA: Weiss S; Boston University Medical Center, Boston, MA: O’Connor G; Kaiser Permanente, Southern California, San Diego, CA: Zeiger R; Washington University of St. Louis, St Louis, MO: Bacharier L; Oregon Health and Science University, Portland, OR: McEvoy C; Indiana University, Riley Hospital for Children: Indianapolis, IN, Tepper R; Pennsylvania State University, University Park, PA: Lyall K; Johns Hopkins Bloomberg School of Public Health Kennedy Krieger Institute, Baltimore, MD:  Landa R; University of California, UC Davis Medical Center Mind Institute, Sacramento, CA:  Ozonoff, S; University of California, UC Davis Medical Center Mind Institute, Davis, CA: Schmidt R; University of Washington: Dager S; 	Children's Hospital of Philadelphia - Center for Autism Research: Schultz R; University of North Carolina at Chapel Hill: Piven J; Johns Hopkins Bloomberg School of Public Health: Volk H; University of Rochester Medical Center Rochester, NY: O’Connor T; University of Pittsburgh Medical Center, Magee Women’s Hospital, Pittsburgh, PA: Simhan H; Harvard Pilgrim Health Care Institute, Boston, MA: Oken E; University of North Carolina, Chapel Hill, NC: O’Shea M, Fry RC; Baystate Children’s Hospital, Springfield, MA : Vaidya R; Beaumont Health Medical Center, Royal Oak, MI: Obeid R; Boston Children’s Hospital, Boston, MA: Rollins C; East Carolina University Brody School of Medicine, Greenville, NC: Bear K; Helen DeVos Children’s Hospital, Grand Rapids, MI: Pastyrnak S; Michigan State University College of Human Medicine, East Lansing, MI: Lenski, M; University of Chicago, Chicago IL: Msall M; University of Massachusetts Medical School, Worcester, MA: Frazier J; Wake Forest Baptist Health (Atrium Health), Winston Salem, NC: Washburn, L; Yale School of Medicine, New Haven, CT: Montgomery A; Michigan State University, East Lansing, MI: Kerver J; Henry Ford Health System, Detroit, MI: Barone, C; Michigan Department of Health and Human Services, Lansing, MI: McKane, P; Michigan State University, East Lansing, MI: Paneth N; Columbia University Medical Center, New York, NY: Herbstman J; University of Illinois, Beckman Institute, Urbana, IL: Schantz S; University of California, San Francisco:, San Francisco, CA: Woodruff T; University of Utah, Salt Lake City, UT: Porucznik CA, Stanford JB, Giardino A; Boston Children’s Hospital, Boston MA: Bosquet-Enlow M; New York University School of Medicine, Karr C; Trasande L; University of California, San Francisco, San Francisco CA: Bush N; University of Minnesota, Minneapolis, MN: Nguyen R; University of Rochester Medical Center: Rochester, NY: Barrett E; University of Wisconsin, Madison, WI: Gern J; Vanderbilt University Medical Center, Nashville, TN: Hartert T; Northeastern University, Boston, MA: Alshawabkeh A;. 

{\it Funding}: The content is solely the responsibility of the authors and does not necessarily represent the official views of the National Institutes of Health. Research reported in this publication was supported by the Environmental influences on Child Health Outcomes (ECHO) program, Office of the Director, National Institutes of Health, under Award Numbers U2COD023375 (Coordinating Center), U24OD023382 (Data Analysis Center), U24OD023319 (PRO Core), UH3OD023251 (Alshawabkeh), UH3OD023320 (Aschner), UH3OD023248 (Dabelea), UH3OD023313 (Deoni), UH3OD023328 (Duarte), UH3OD023318 (Dunlop), UH3OD023279 (Elliott), UH3OD023289 (Ferrara/Croen), UH3OD023282 (Gern), UH3OD023365 (Hertz-Picciotto), UH3OD023244 (Hipwell), UH3OD023275 (Karagas), UH3OD023271 (Karr), UH3OD023347 (Lester), UH3OD023389 (Leve), UH3OD023268 (Weiss), UH3OD023288 (McEvoy), UH3OD023342 (Lyall), UH3OD023349 (O’Connor), UH3OD023286 (Oken), UH3OD023348 (O’Shea), UH3OD023285 (Kerver), UH3OD023290 (Herbstman), UH3OD023272 (Schantz), UH3OD023249 (Stanford), UH3OD023337 (Wright), UH3OD023305 (Trasande).  
{\it Conflict of Interest}: None declared.

\bibliographystyle{biorefs}
\bibliography{samplebibtex}

\end{document}